Category: Human Information Interaction
Title: **The Investigation of Social Media Data Thresholds for Opinion Formation**

Asher, D.[1], Caylor, J.[1], Mittrick, M.[1], Richardson, J.[1], Heilman E.,[1] Bowman, E.[1], Korniss, G.[2], Szymanski, B.[2]

[1] Army Research Laboratory (ARL), Aberdeen Proving Ground, MD
[2] Rensselaer Polytechnic Institute (RPI), Troy, NY

This research was supported in part by the Army Research Laboratory under Cooperative Agreement Number W911NF-09-2-0053 (the ARL Network Science CTA). The views and conclusions contained in this document are those of the authors and should not be interpreted as representing the official policies either expressed or implied of the Army Research Laboratory or the U.S. Government.

# Abstract

The pervasive use of social media has grown to over two billion users to date, and is commonly utilized as a means to share information and shape world events. Evidence suggests that passive social media usage (i.e., viewing without taking action) has an impact on the user's perspective. This empirical influence over perspective could have significant impact on social events. Therefore, it is important to understand how social media contributes to the formation of an individual's perspective. A set of experimental tasks were designed to investigate empirically derived thresholds for opinion formation as a result of passive interactions with different social media data types (i.e., videos, images, and messages). With a better understanding of how humans passively interact with social media information, a paradigm can be developed that allows the exploitation of this interaction and plays a significant role in future military plans and operations.

# 1. Introduction

Social media viewing or passive social media consumption has been shown to shape an individual's perspective (i.e., opinion) [1]. Given the influence social media can have in shaping the individual's opinion, it is important to understand opinion formation, the factors that contribute to the shifts in opinion, and the dynamics associated with the formation and shifting of opinions. Much theoretical work has focused on opinion dynamics through the use of computational models and simulation experiments [2]–[6], [6]–[10]. Empirical studies have investigated opinion polarization [11] and opinion evolution [12]; however, empirical research is still needed to understand how passive social media consumption of different data types (e.g., images, videos, and messages of pure text) contributes to opinion changes. Specifically, in exploring how people form and change opinions, the various forms of bias are critical considerations.

Category: Human Information Interaction
Title: **The Investigation of Social Media Data Thresholds for Opinion Formation**

Content bias has been extensively investigated within the context of media [13]–[17], and is in line with the findings associated with social media content [12]. The evidence suggests that *a priori* perspective and personal experience result in polarization from the consumption of content [12]. From this evidence, it would be of interest to understand how opinions form without the built-in bias of experience related to the content. Our approach aims to investigate how opinions form without the bias of content.

Information media sources are pervasive within most areas where military operations occur. Military information operation analysts manipulate and control the information environment to provide commanders with a decisive advantage over adversaries, threats, and enemies [18]. Determining the types and number of media to present to a populous with intentions of altering opinions is highly relevant to military information operations. The results from this experiment represent a first step toward identifying the effort needed to achieve opinion formation in a given population. Alternatively, commanders are interested in understanding existing opinions held by individuals and social groups within an area of operations to develop appropriate courses of action.

Social influence has been shown to have a strong effect over the formation of an individual's opinion [19], [20]. The term can be interpreted as conformity, peer pressure, and compliance resulting from the expressed opinions or perspectives of other individuals. Within the context of this paper, social influence is meant to capture the contribution that a like-minded group makes to opinion formation versus a group holding an opposing perspective.

Context provides a framework for pieces of information, essentially bonding these pieces into a set and resulting in reduced variance over the interpretation and increased coherence of the information as a whole. In the absence of context, pieces of information are subject to individual interpretation, resulting in increased variance of interpretation and a potential general lack of coherence over the whole set. In the study, we address context using levels of controversy. Different levels of controversy helped subjects estimate how pieces of information contribute to the formation of an opinion.

Crowdsourcing is a term used to describe the use of crowds to answer difficult questions or solve difficult problems [21]. This phenomena is based on the Wisdom of the Crowd concept, in that a large group of aggregated answers has generally been found to be as good as or often better than a single answer by any person in the group [22]. Recently, the method of crowdsourcing was used to help researchers discover the most likely folding patterns for proteins by turning the problem into an online game called Foldit [23], [24]. Given the power of the crowdsourcing method, it is an ideal choice for estimating the number of different data-types that lead to an individual's formation of an opinion.

Category: Human Information Interaction
Title: **The Investigation of Social Media Data Thresholds for Opinion Formation**

The efficacy of crowdsourcing comes from the concept "wisdom of the crowd," which indicates that a large number of individuals estimating some phenomena will produce an averaged estimate that is as good as or often better than that of an expert [25]. An explanation for this phenomenon is that noise inherently exists in estimates and an average over a large amount of these noisy estimates results in a reduction in the overall noise, abiding by the law of large numbers in probability theory [26]. This makes crowdsourcing an ideal method for estimating empirical thresholds of passive social media consumption for opinion formation.

The aim of this study was to illuminate opinion formation thresholds from passive social media consumption. To eliminate content bias while maintaining information coherence, a general and generic content-free framework with differing contexts (represented by levels of controversy) was established. To the best of our knowledge, no prior work has examined how passive social media data consumption contributes to opinion formation in the absence of content. Furthermore, we used levels of controversy and social influence to understand how opinion formation depends on these factors.

## 2. Methods

Amazon Mechanical Turk (MTurk) was used to collect the human subject data from 235 participants. A simple computerized task required subjects to enter numbers in boxes that represented their estimates of the amount of data-types expected to be viewed in a static timeframe (one day) before formulating an opinion, given a specified context. 235 subjects were randomly assigned to 1 of 28 conditions. A condition consisted of one data-type presentation within one context and three questions related to information source embedded in.

### Human Subjects

The study falls under the Army Research Laboratory (ARL) internal review board (IRB) Exempt Research Determination for Protocol ARL 17-087, which exempts the study from regulation 32 CFR 219. The research falls into the exemption criteria defined by the Common Rule, which states that human subjects cannot be identified by the collected data; and the responses provided by the subjects place them at no risk of criminal or civil liability, nor could they be damaging to their financial standing, employment, or reputation.

Upon selecting to participate in the study for a quarter (25¢), subjects were notified that it would require approximately three minutes to complete, and no personally identifiable information would be collected. Subjects needed to complete all questions in order to be compensated. 12 demographic questions focused primarily on social media usage, and three task-related questions established the data per participant. Exclusionary criteria consisted of

Category: Human Information Interaction
Title: **The Investigation of Social Media Data Thresholds for Opinion Formation**

the subjects' general use of social media. If a participant indicated that they did not use social media, they were thanked for their interest in the study and their participation was ended without collecting any data.

User bias was minimized by allowing each subject to participate in the study only once. The MTurk account name was used solely to determine if a subject had attempted to participate previously. If a subject attempted to participate a second time the program informed them that they were no longer eligible.

### Data-types

Subjects were asked to estimate their opinion formation thresholds for three distinct data-types: 1) *Images*, 2) *Videos*, and 3) *Messages*. These data-types were selected for their easily identifiable differences. Subjects were shown the following descriptions corresponding to the data-types:

*Messages* –data-type includes text, a tweet, or a post on Facebook.
*Videos* –data-type includes any moving pictures, animations, GIFs, and videos.
*Images* –data-type includes still pictures, images, and drawings.

### Data-type Orientations

To account for potential data-type interactions, we tested seven arrangements or presentations of the data-types (see Figure 1). Our general hypothesis regarding data-type interactions is that the population of subjects will not have significantly adjusted their estimates for the different data-type presentations. This suggests that the estimated threshold for opinion formation from data-type presentation I (*Images* alone) should not be significantly different from the estimates for Images in presentations IV, V, or VII (see Figure 1). Therefore, the different data-type presentations were selected to provide sufficient evidence for only utilizing a single presentation in future studies (i.e., data-type presentation VII, see Figure 1).

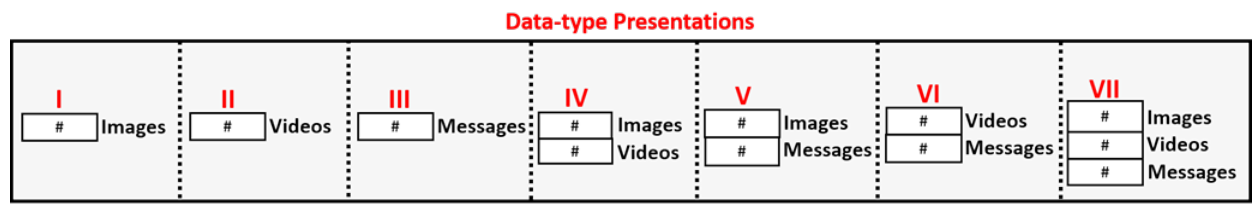

**Figure 1. Data-type presentations:** The different orientations of the representative data-types (*Images*, *Videos*, and *Messages*) were used in the study to determine if data-type interactions needed to be considered for opinion formation threshold estimates.

Category: Human Information Interaction
Title: **The Investigation of Social Media Data Thresholds for Opinion Formation**

## Contexts

In order to counter-balance for the abstraction introduced by excluding content from the experiment, context was introduced as levels of controversy. For the purposes of this article, the "controversy" and "context" are used interchangeably. The four levels of controversy (*None*, *Low*, *Medium*, and *High*) were used to investigate influence over opinion formation thresholds. To summarize, the four contexts were:

*None* – no reference to controversy
*Low* – minimal controversy (some people would have an opinion)
*Medium* – controversial (many people would have an opinion)
*High* – highly controversial (most or all people would have an opinion)

With an exception for the '*None*' case, the different levels of controversy were introduced to the subjects with an example and a color coded word (see Figure 2). These levels were selected to try and shift the opinion formation threshold estimates of the subjects, essentially allowing us to observe how different contexts influence opinion formation thresholds.

> NONE → no context referenced
>
> LOW → an example of a LOW level of controversy is:
> A car company introduces a new standard car color in hot pink.
>
> MEDIUM → an example of a MEDIUM level of controversy is:
> A typically conservative state (e.g., Texas) approves a liberal law (e.g., recreational marijuana).
>
> HIGH → an example of a HIGH level of controversy is:
> A dictator-run country (e.g., North Korea) fires a chemical weapon into a US allied country (e.g., France).

**Figure 2. Contexts:** Four different contexts (*None*, *Low*, *Medium*, and *High*) with three different questions. Subjects were provided with tangible examples to help introduce context into the experiment.

## Data Collection

Each MTurk subject was randomly assigned one condition and asked to answer three questions investigating the influence of different sources. This implies that each condition has independent data. The only dependencies between the data are across the three questions a subject answered. The experiment contained seven data-type presentations (see Figure 1) and

Category: Human Information Interaction
Title: **The Investigation of Social Media Data Thresholds for Opinion Formation**

four contexts, which resulted in 28 conditions distributed over 235 subjects participating in the study (see Figure 3).

The three questions were used to measure the influence of the source of the social media data-type being estimated and included the following sources:

*Question 1*: Before you FORM an OPINION how many data types listed below would you expect to view in a day?
*Question 2*: Before you FORM an OPINION how many data types listed below would you expect to view in a day, given that the data type(s) were posted by people who think like you?
*Question 3*: Before you FORM an OPINION how many data types listed below would you expect to view in a day, given that the data type(s) were posted by people with different viewpoints?

The first question did not specify a source of the social media posts, and it were used as a control case. The second question emphasized that the social media data were posted by like-minded people, aiming to measure the influence that in-group posts have on a subject's estimate of opinion formation threshold. The third question emphasized that the social media data were posted by multiple groups with different perspectives, capturing the combination of in-group and out-group influence. Together, the three questions allow us to measure influence from different sources over opinion formation threshold.

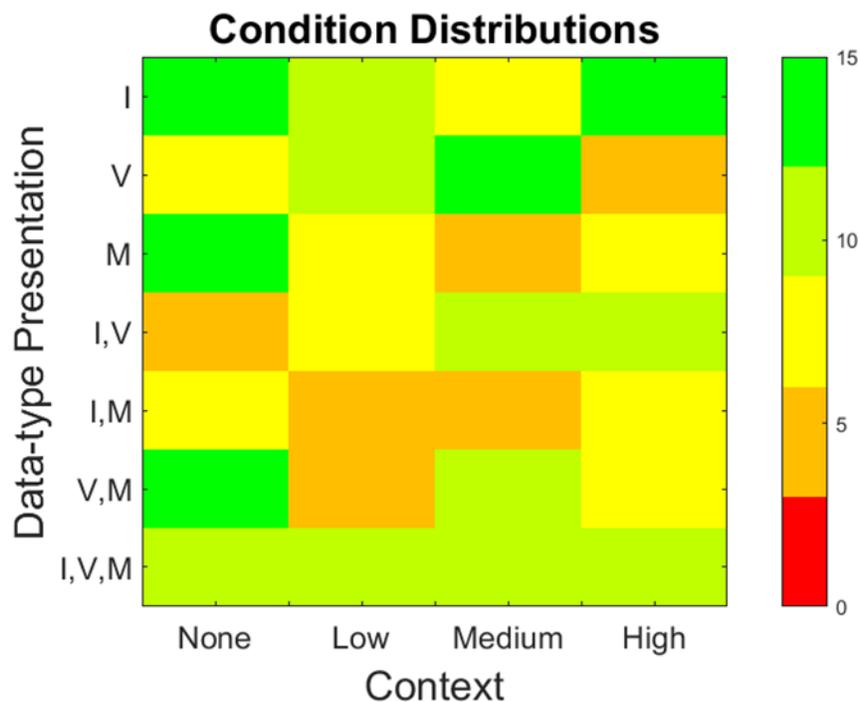

Category: Human Information Interaction
Title: **The Investigation of Social Media Data Thresholds for Opinion Formation**

**Figure 3. Condition distributions:** The image shows the distribution of 235 subjects across 28 conditions. The y-axis represents the seven distinct data-type presentations (see Figure 1) with I -> *Images*, V -> *Videos*, and M -> *Messages*. The x-axis shows the different contexts (see Figure 2). The color legend shows the number of subjects corresponding to the binned color group. The figure illuminates the number of subjects per condition.

## Data Analysis Techniques

Non-parametric tests are needed if the data are not distributed normally. The Jarque-Bera (JB) goodness-of-fit test indicates if a data sample came from an unspecified normal distribution. Therefore, the JB test was used to determine if the data from the respective conditions, questions, data-type presentations, and data-types were normally distributed.

The Wilcoxon rank-sum test or the Mann-Whitney U test is a non-parametric test. The null hypothesis we used here for the Wilcoxon rank-sum test states that the distributions of the compared samples are equal. Small p-values indicate that the null should be rejected and the distributions are not equal. The two compared samples are assumed to come from identical and continuous distributions with a possible shift.

To quantify differences between the conditions, the Kruskal-Wallis test was used. This is a non-parametric method similar to a classical one-way ANOVA. The Kruskal-Wallis test is equivalent to the Wilcoxon rank sum test for equal medians but allows comparisons with more than two groups. Medians of the subject data from the different conditions were compared to determine if the samples are from the same underlying distribution. This test assumes that all samples come from populations having the same continuous distribution and that all observations are mutually independent. In comparison to the Kruskal-Wallis test, a classical one-way ANOVA replaces the "same continuous distribution" assumption with a stronger assumption that the populations are normally distributed. However, in the study, normally distributed data were not observed.

The Kruskal-Wallis test uses ranks of the subject data, rather than numeric values, to compute the test statistics. It finds ranks by ordering the data from smallest to largest across all compared groups, taking the numeric index of this ordering. The rank for a tied observation is equal to the average rank of all observations tied with it. The F-statistic used in classical one-way ANOVA is replaced by a chi-square statistic, and the p-value measures the significance of the chi-square statistic.

The p-values generated from the Kruskal-Wallis test represent the statistical significance associated with all compared data from the respective conditions originating from the same distribution (i.e., the null hypothesis). P-values are considered significant if below 0.01 and corrections for multiple comparisons were not necessary.

Category: Human Information Interaction
Title: **The Investigation of Social Media Data Thresholds for Opinion Formation**

# 3. Results

To assess the factors contributing to opinion formation estimates across the different conditions, the analyses necessitated the segmentation of data by data-type (see Section 2 Methods: Data-types). With this segmentation, data were measured under four data-type presentations. For example, the data-type *Images* (I), was compared to *Images and Videos* (I, V), *Images and Messages* (I, M), and *Images and Videos and Messages* (I, V, M). However, the analysis of data-type presentations for *Images* did not include *Videos* (V), *Messages* (M), or *Videos and Messages* (V, M). In contrast, all three data-types were compared across the four contexts (see Section 2 Methods: Contexts) and three questions (see Section 2 Methods: Data Collection).

## Jarque-Bera Tests

The JB test for normally distributed data revealed that less than 50% of the data per data-type were normally distributed (46% *Images*; 46% *Videos*; 44% *Messages*), indicating that non-parametric tests such as Kruskal-Wallis and Wilcoxon rank sum tests are appropriate for further analysis.

## Wilcoxon Rank Sum Tests

Pair-wise Wilcoxon rank sum tests were performed between each data-type presentation (six comparisons) per context, question, and data-type. The Bonferroni corrected significance level alpha initially set to 0.01 was corrected to 0.0017. At this significance level, no pairwise compared medians reached significance when correcting for multiple comparisons. This result indicates that thresholds across the four data-type presentations per data-type (e.g., data-type: *Images* => four data-type presentations: 1. **I**; 2. **I**, V; 3. **I**, M; 4. **I**, V, M) were not statistically different. Therefore, the medians (i.e., opinion formation thresholds) were not found to be significantly different across the data-type presentations within a data-type, context, and question (e.g., data-type: *Videos*, context: *Low*, question: Q1).

## Kruskal-Wallis Tests

The Kruskal-Wallis tests were first conducted across the three questions. This indicates that the subject responses for a data-type presentation (e.g., I, V) and a context (e.g., *Medium*) were compared across the three questions. The results suggest that the Kruskal-Wallis test for the same underlying distribution was not significant when comparing across contexts (see Figure 4A – 4C). This indicates that we cannot conclude that the samples are from the same distribution, however, it does not indicate that the samples are significantly different.

Category: Human Information Interaction
Title: **The Investigation of Social Media Data Thresholds for Opinion Formation**

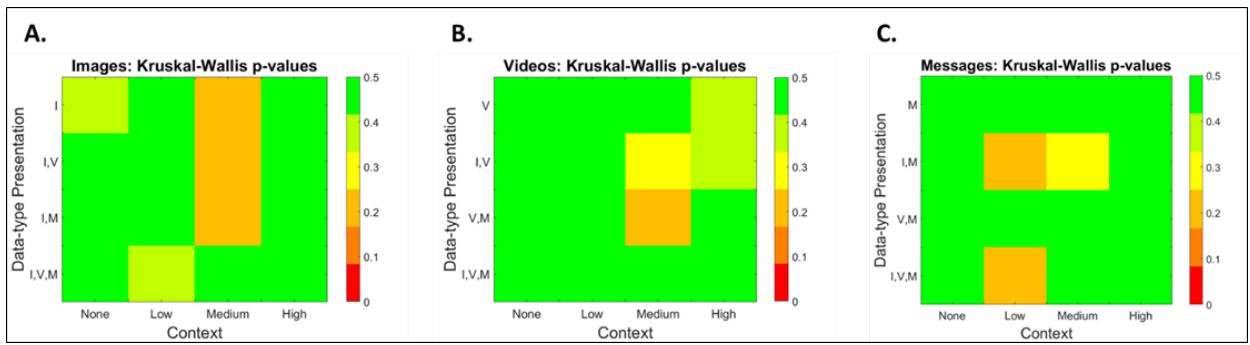

**Figure 4. Kruskal-Wallis p-values across questions:** The graphs show p-values calculated from Kruskal-Wallis tests for null hypothesis (same underlying distribution) across questions, per data-type presentation (y-axis) and per context (x-axis). There are a total of 16 p-values were evaluated corresponding to the four data-type presentations by the four contexts. The color legends show the p-values, truncated at 0.50. All p-values for A, B, and C were above 0.01, which indicates that the questions did not result in the same underlying distribution for any of the data-type presentation/context comparisons. A. *Images*, B. *Videos*, C. *Messages*.

The p-value comparisons shown in Figure 4 clearly illustrate the absence of a significant null hypothesis, which states that the compared samples come from the same underlying distribution. A significant null result would indicate that the three questions did not result in different underlying distributions, suggesting that the questions may have had a significant influence over the subjects' opinion formation threshold estimates.

Although none of the comparisons in Figure 5 reached significance, six of the comparisons did reach marginal significance (see red-colored comparisons in Figures 5B and 5C). For the data-type *Videos,* the p-values across contexts show that one data-type presentation (I, V) reached marginal significance (see Figure 5B) for both no source specified (p = 0.0345; *Question 1*) and multiple different perspectives (p = 0.0217; *Question 3*). Similarly, the data-type *Messages* showed four marginally significant p-values at data-type presentation: M and question: No Source (p = 0.0356), data-type presentation: M and question: Different (p = 0.0184), data-type presentation: V, M, and question: Like-Minded (p = 0.0700), and data-type presentation: V, M, and question: Different (p = 0.0225). Given that none of the p-values were < 0.01, these results suggest that context may not have an influence over opinion formation thresholds in the aforementioned cases.

Category: Human Information Interaction
Title: **The Investigation of Social Media Data Thresholds for Opinion Formation**

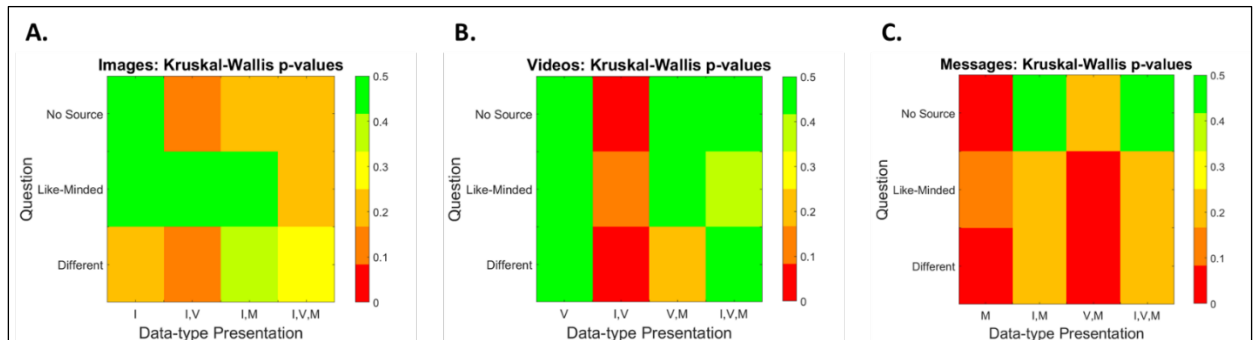

 **Figure 5. Kruskal-Wallis p-values across contexts:** The graphs show p-values calculated from Kruskal-Wallis tests for null hypothesis (same underlying distribution) across contexts, per question (y-axis) and per data-type presentation (x-axis). A total of 12 p-values were computed corresponding to the three questions by the four data-type presentations. The color legends show the p-values, truncated at 0.50. All p-values for A, B, and C were above 0.01, which indicates that the contexts did not result in the same underlying distribution for any of the question/data-type presentation comparisons. A. *Images*, B. *Videos*, C. *Messages*.

Together, the Kruskal-Wallis tests showed that the opinion formation thresholds may have been influenced by source (*Questions 1–3*) and context (see Figure 2). In addition, several cases reached marginal significance (see Figures 5B and 5C). To better understand the marginally significant cases, additional data could help, given that the number of subjects per condition had large differences (see Figure 3).

## Opinion Formation Threshold Estimates

Due to the uneven number of data points per condition (see Figure 3) and the lack of normally distributed data, medians per data-type (*Images*, *Videos*, and *Messages*) from the respective conditions were used to determine the opinion formation thresholds. Although possible, no calculated median was zero and across all data-types, contexts, questions, and data-type presentations the range is [1, 54]. Overall, the *Images* data-type tended to have higher threshold values compared to *Videos and Messages* (compare Figure 6A – 6C to Figures 7A – 7C and 8A – 8C). This result indicates that subjects estimated that a larger number of images would be needed to form an opinion.

Category: Human Information Interaction
Title: **The Investigation of Social Media Data Thresholds for Opinion Formation**

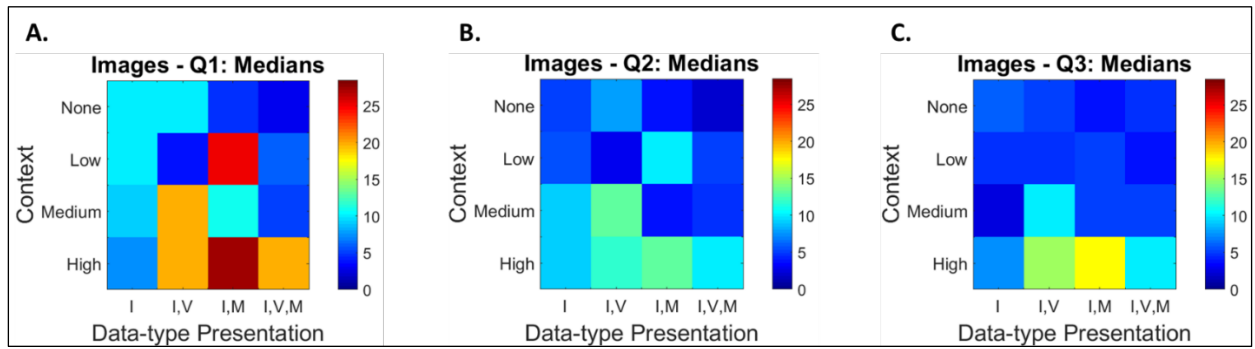

**Figure 6.** *Images* **data-type medians:** The figures show medians calculated from the varying number of subjects per condition (see Figure 3). The y-axes represent the four different contexts, the x-axes represent the four relevant data-type presentations, and the color bars show the full range of median values for the *Images* data-type across the three questions. (See Section 2 Methods: Data Collection).

Select trends can be observed in the threshold data for the *Images* data-type. For example, the context *High* had a clear increase in median values (compare *High* row to other contexts in Figure 6A – 6C). This result suggests that the context with high controversy shifted the opinion formation thresholds up when compared to the other contexts. Although these values are higher, more data are needed to confirm statistical significance. Similarly, the data in Figure 6 show that the data-type presentations I, V and I, M appear to have overall higher thresholds when comparing across contexts and questions. Finally, the thresholds tend to decrease when comparing across questions (compare Figure 6A to 6B and 6B to 6C). These results, though not necessarily significantly different, show trends that may become significant with additional data.

Surprisingly, some of the trends for the *Videos* data-type are similar to the *Images* data-type. The data show that the context *High* has greater threshold values compared to other contexts within a question (compare Figure 7 *High* rows to other three contexts), and the context *High* decreases across questions (compare Figure 7A to 7B and 7B to 7C). Notably, the maximum median for the *Videos* data-type is 10, indicating that subjects generally estimate lower opinion formation thresholds for the *Videos* data-type versus the *Images* data-type (compare Figure 6 to 7). Furthermore, this result is intuitive since a person would generally expect a video to take more time to view than an image, which implies that someone can see more images in the same time they would watch fewer videos.

Category: Human Information Interaction
Title: **The Investigation of Social Media Data Thresholds for Opinion Formation**

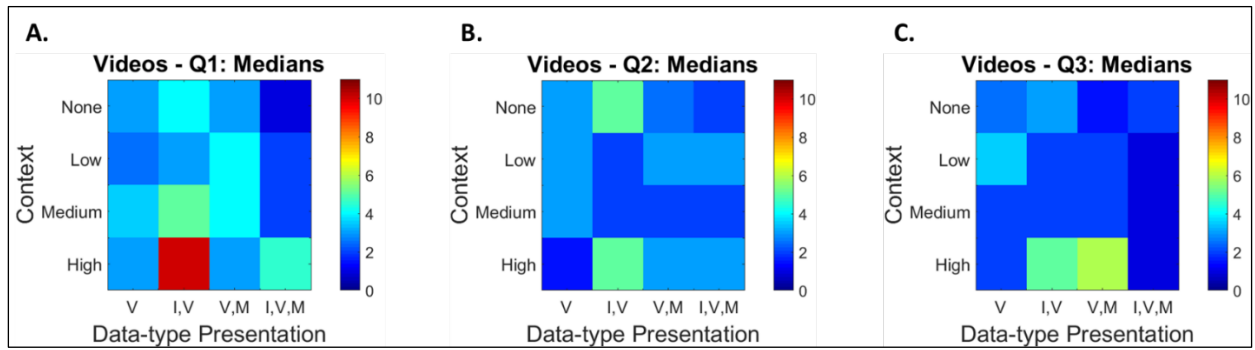

**Figure 7. *Videos* data-type medians:** The figures show medians calculated from the varying number of subjects per condition (see Figure 3). The y-axes represent the four different contexts, the x-axes represent the four relevant data-type presentations, and the color bars show the full range of median values for the *Videos* data-type across the three questions. (See Section 2 Methods: Data Collection).

In contrast to both *Videos* and *Images* data-types, the *Messages* data-type shows different trends. The *Messages* data-type shows consistencies between questions 1 and 3 (compare Figure 8A to 8C), whereas question 2 appears to have a different pattern (compare Figure 8B to 8A and 8C). In addition, the threshold for context Low and data-type presentation V, M is approximately five times higher than all other conditions. This can be explained by a small number of subjects (n = 4), and by the fact that half the data points could potentially be outliers. The trend persists across the questions because subjects were asked to provide estimates to all three questions, which continued to capture their potentially unreasonable opinion threshold estimates. As with the other data-types, additional data are likely to produce results without outliers.

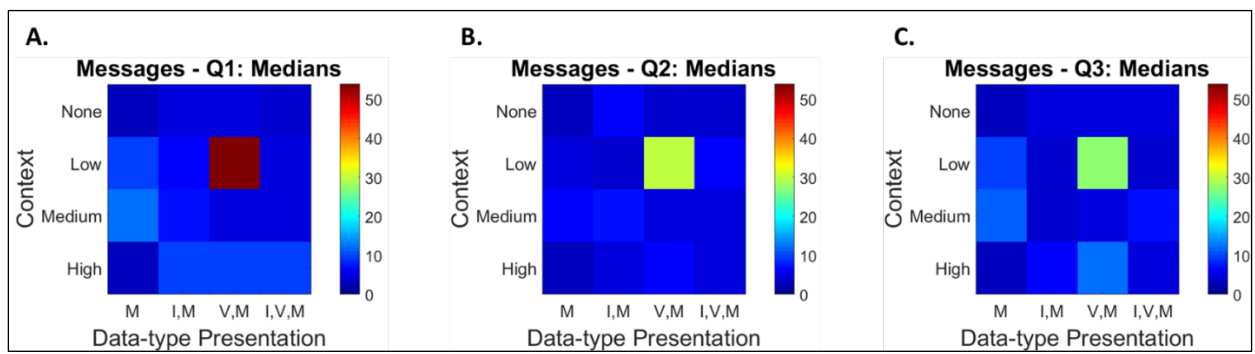

**Figure 8. Messages data-type medians:** The figures show medians calculated from the varying number of subjects per condition (see Figure 3). The y-axes represent the four different contexts, the x-axes represent the four relevant data-type presentations, and the color bars show the full range of median values for the *Messages* data-type across the three questions. (See Section 2 Methods: Data Collection).

Together, the medians show tentative opinion formation thresholds derived from the distributions of subject responses to the various conditions. The results from the Wilcoxon

Category: Human Information Interaction
Title: **The Investigation of Social Media Data Thresholds for Opinion Formation**rank-sum tests indicate that the data-type presentations did not have a significant effect on subjects' opinion formation thresholds. However, additional data will be collected to resolve potential outliers within the respective distributions and illustrate potential significant differences between opinion formation thresholds.

# 4. Discussion

The aim of this study was to illuminate opinion formation thresholds from passive social media consumption. To eliminate content bias while maintaining information coherence, a general and generic content-free framework with differing contexts was established. To the best of our knowledge, no prior work has examined how passive social media data consumption contributes to opinion formation thresholds in the absence of content. Furthermore, we used levels of controversy (see Figure 2) and social influence (see Section 2 Methods: Data Collection) to understand how opinion formation depends on these factors.

Evidence suggests that passive social media usage (i.e., viewing without taking action) has an impact on the user's perspective [1]. This influence over the individual's perspective could have a significant impact on social events. For this reason, an experimental task was developed to determine thresholds for opinion formation from passive interactions with different social media data-types (i.e., *Images*, *Videos,* and *Messages*). The results suggest 1) different presentations of the data-types (see Figure 1) did not have a significant effect over opinion formation threshold (see Section 3 Results: Wilcoxon Rank Sum Tests), 2) contexts and source information (*Questions 1–3*) appear to have a differential impact on opinion formation (see Figures 4 and 5), and 3) opinion formation thresholds do appear to exist, and are data-type dependent (see Figures 6–8). More data are needed to substantiate conclusions made for results 2) and 3).

Crowdsourcing is an ideal method for understanding phenomena that is inherently noisy like opinion formation threshold estimates. In our experimental paradigm, 235 subjects participated in one of 28 conditions, leaving several conditions relatively underrepresented (see Figure 3). This underrepresentation was primarily due to randomizing the condition selection criteria and the lack of subjects. In light of this, the 235 subjects were sufficient to test the impact of data-type presentations (see Figure 1), finding them non-significant (see Results section: Wilcoxon Rank Sum Tests). This implies that future experiments regarding opinion thresholds will not require different data-type presentations.

Some work has been done to understand how small groups of individuals' opinions can influence public opinion [27], but little has focused upon the factors that drive an individual toward the formation of an opinion based solely on passive social media consumption. Given the novelty of this work, it is reasonable to remain skeptical about the existence of opinion formation thresholds without additional experimentation. However, the results from this experiment suggest that these opinion formation thresholds do exist and are independent of data-type presentation but dependent upon the data-type. Additional data will need to be collected to confirm these results.

Category: Human Information Interaction
Title: **The Investigation of Social Media Data Thresholds for Opinion Formation**